\documentclass[a4paper,10pt,onecolumn,final,accepted=2026-04-09]{quantumarticle}
\pdfoutput=1
\usepackage{amsmath}
\usepackage{amssymb}
\usepackage{graphicx}
\usepackage{dcolumn}
\usepackage{bm}
\usepackage[colorlinks = true, allcolors = blue]{hyperref}
\usepackage{verbatim}
\usepackage{soul}
\usepackage{float}
\usepackage{xcolor}
\usepackage{multirow}
\usepackage{orcidlink}
\usepackage{animate}
\usepackage[numbers,sort]{natbib}

\newenvironment{ruledtabular}
{\vspace{2mm}\hrule\vspace{1mm}
	\centering
	\begin{minipage}{\textwidth}}
	{\end{minipage}
	\vspace{1mm}\hrule\vspace{2mm}}
\begin{document}
\title{Quantum stick-slip motion in nanoscaled friction}

\author{Dai-Nam Le\,\orcidlink{0000-0003-0756-8742}}
\email{dainamle@usf.edu}
\thanks{Homepage: \href{https://sites.google.com/view/dai-nam-le/}{https://sites.google.com/view/dai-nam-le/}}
\affiliation{Department of Physics, University of South Florida, Tampa, Florida 33620, USA}

\author{Pablo Rodriguez-Lopez\,\orcidlink{0000-0003-0625-2682}}
\email{pablo.ropez@urjc.es}
\affiliation{{\'A}rea de Electromagnetismo and Grupo Interdisciplinar de Sistemas Complejos (GISC), Universidad Rey Juan Carlos, 28933, M{\'o}stoles, Madrid, Spain}

\author{Lilia M. Woods\,\orcidlink{0000-0002-9872-1847}} 
\email{lmwoods@usf.edu}
\thanks{Corresponding authors; Homepage: \href{https://www.amd-woods-group.com/}{https://www.amd-woods-group.com/}}
\affiliation{Department of Physics, University of South Florida, Tampa, Florida 33620, USA}

\begin{abstract}
Friction in atomistic systems is usually described by the classical Prandtl-Tomlinson model suitable for capturing the dragging force of a nanoparticle in a periodic potential. Here we consider the quantum mechanical version of this model in which the dissipation is facilitated by releasing heat to an external bath reservoir. The time evolution of the system is captured with the Liouville-von Neumann equation through the density matrix of the system in the Markov approximation. We examine several kinetic and dissipative properties of the nanoparticle by delineating classical vs quantum mechanical effects. We find that the Landau-Zener tunneling is a key factor in the overall reduction of the frictional dissipation when compared to the classical motion in which such tunneling is absent. {Other regimes of motion, controlled by the corrugation parameter and other properties, are also found }. This in-depth study analyzes the interplay between velocity, strength of interaction, and temperature to control the frictional {force} and provide useful guidelines for experimental data interpretation. 

\end{abstract}

\maketitle

\section{Introduction}
The relative motion between objects in close contact causes friction as a result of energy dissipated in this dynamical process. Friction often leads to reduced efficiency and reliability of machines and devices. Although friction has been extensively studied, the governing mechanisms are not completely understood yet \cite{Volokitin2007,Vakis2018, Decca2020, Lombardo2021, Modi2026}. Computational studies using first principles methods have examined adhesion, corrugation, and van der Waals interaction energies for various materials \cite{Wolloch2019,Cahangirov2012, Restuccia2023, Losi2023, Woods2024, Silvestrelli2025, Dang2025}. These directly related to static friction properties are instrumental in identifying  materials with targeted tribological applications \cite{Wolloch2018,Torche2022}. 

The dynamical side of frictional processes is important in force microscopy applications \cite{Szlufarska2008} and typically, it is studied with continuum approaches that are variations of the Prandtl–Tomlinson model \cite{Tomlinson1929,Prandtl1928}. This is essentially a classical approach in which a moving particle without internal degrees of freedom experiences stick-slip motion above a molecular chain represented as an oscillatory-like periodic potential. The dynamics of the process is modeled semi-empirically with materials-dependent parameters for the corrugation, velocity, and particle-chain interaction. The dynamical frictional properties of atomic/molecular chains in relative motion have also been studied within the Frenkel-Kontorova model \cite{Krajewski2004,Krajewski2005,Xu2007}, which was also used to understand dislocations, nonlinear effects and topological  defects in relative chain displacements. Numerical methods, such as path integrals and quantum Monte Carlo simulations have shown the role of quantum and thermal fluctuations of the forced motion of an atom above a finite discrete chain. However, many-body collective phenomena at the quantum mechanical level are difficult to access via such algorithms due to numerical limitations associated with the size of the studied system.

A basic quantum mechanical effect important for the dynamics of frictional motion is the mixing between the energy states of the system creating conditions for tunneling. This is captured by the most basic assumptions in the Landau-Zener (LZ) theory \cite{Landau1932,Zener1932} in which the energy difference between two adiabatic states is violated by creating avoided level crossings. The LZ theory has found applications in dissipative and multilevel systems \cite{Pokrovsky2002,Huang2018,Zanca2018}. Curiously, a dissipative LZ two level system coupled to a bath of harmonic oscillator shows that the environmental temperature can actually enhance the probability of the system remaining in its ground state leading to quantum annealing \cite{Arceci2017}.

In this paper, we investigate the quantum mechanical  Prandtl–Tomlinson model in the presence of an external bath reservoir into which heat is released in the frictional dissipation. {This problem was considered in Ref. \cite{Zanca2018} where frictional quantum lubricity was found in the frictional dissipation rate at the adiabatic regime. Here  we define  several  energetic, kinetic, and dissipative properties including average velocity, power, released heat, frictional force, and geometric phase. These properties are examined by solving the Liouville-von Neumann equation in the Markov approximation beyond the adiabatic regime.} To better understand the quantum mechanical nature of the motion, we also investigate the quantum mechanical Prandtl–Tomlinson model in the absence of an external bath as well as the classical limit with and without an external bath using the stochastic Newton law. {The LZ tunneling is a key mechanism for quantum lubricity in stick-slip motion (as also found in Ref. \cite{Zanca2018}), however, we find that nanoscaled friction may exhibit other types of motion depending on the physical parameters in the Prandtl–Tomlinson model. Furthermore, the newly analyzed} frictional force and released heat, in particular, give a quantitative representation of the frictional process and can be directly associated with experimental measurements.

\section{Model system}

\begin{figure}[H]
    \centering
    \includegraphics[width = 0.6 \textwidth]{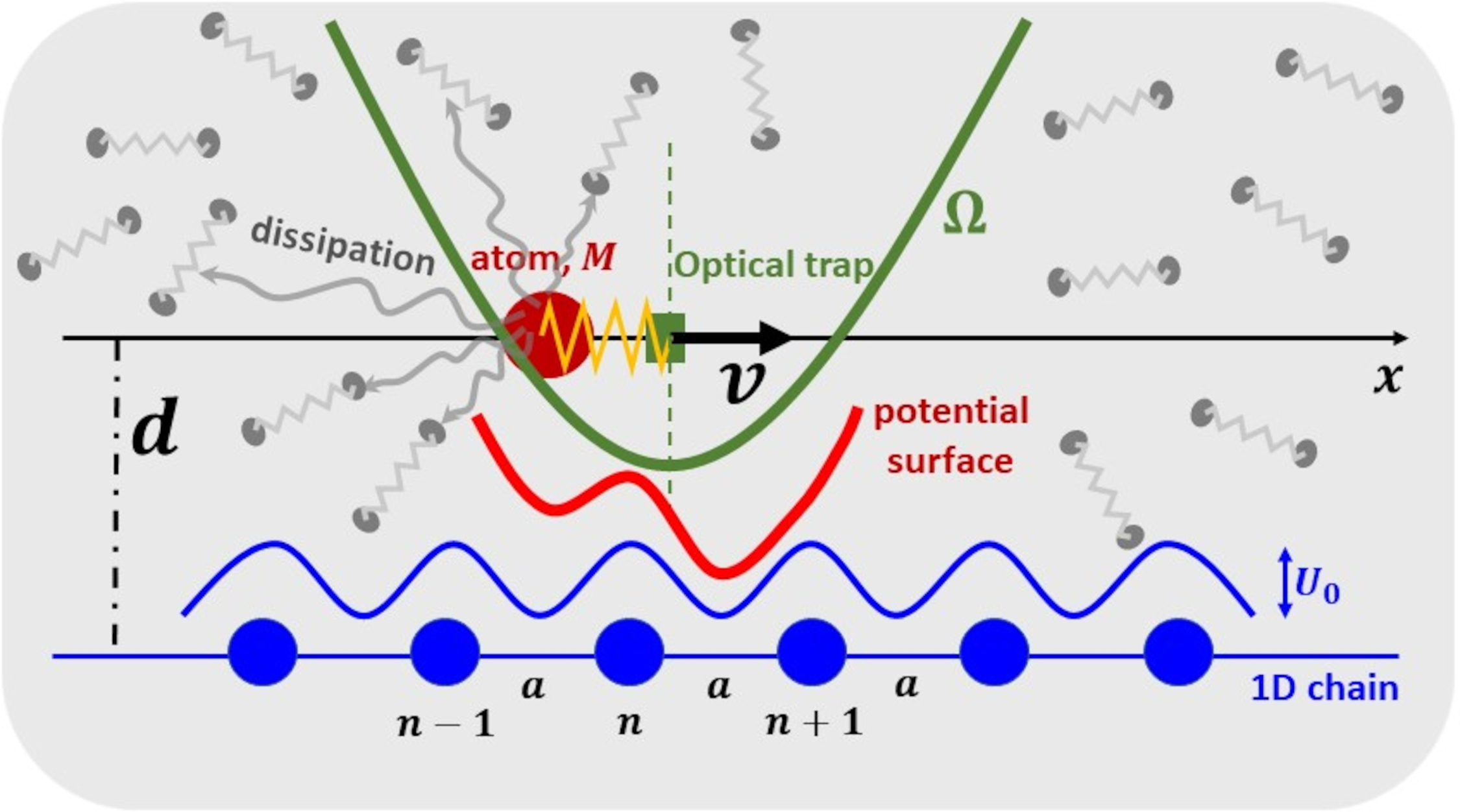}
    \caption{\label{fig:1} Schematics of a particle with mass $M$ confined in a harmonic potential moving with a constant velocity $v$ above an infinitely long atomic chain with periodicity $a$. The distance $d$ between the particle and the chain is maintained throughout the motion. 
    }
\end{figure}

The system under consideration consists of a {nanoscale} particle with mass $M$ at a distance $d$ above a one-dimensional infinitely long atomic chain with lattice period $a$, as shown in Fig. \ref{fig:1}.  The particle moves with a constant velocity $v$ parallel to the chain. To maintain this motion, it is assumed that $M$ is in an optical trap  modeled by a harmonic potential. The particle also experiences the influence of a short-ranged interaction from the overlap of its atomic orbitals and the atomic orbitals of the chain. Additionally, there is a long-ranged van der Waals coupling originating from the exchange of electromagnetic fluctuations between $M$ and the chain. The optical trap balances these different interactions by maintaining a motion parallel to the chain at a constant velocity.

This system can be modeled by a 1D Hamiltonian given as 
\begin{equation}
    \label{eqn:HQ}
    \hat{H}_S (t) = \frac{1}{2M} \hat{p}^2 + \frac{1}{2} M \Omega^2 \left( x - v t \right)^2 + U_0(d) \sin^2 \left( \frac{\pi}{a} x \right) + \Delta_0 (d) ,
\end{equation}
where the first term denotes the kinetic energy of the particle with its momentum operator $\hat{p}$. The second term contains a time-dependent harmonic term with a spring constant $M \Omega^2$ associated with the moving optical trap of the nanoparticle ($\Omega$ is a characteristic frequency). The spatial periodic term and the constant $\Delta_0 (d)$ reflect the combined effect of short-ranged interatomic and long-ranged van der Waals interactions. Assuming that $d \gg a$, one finds that $U_0(d)=A_s\sqrt{\frac{2\pi d d_s}{a^2}}e^{-\frac{d}{d_s}-\frac{2\pi^2 dd_c}{a^2}}-\frac{3\pi C_6}{8ad^2}e^{-\frac{2\pi d}{a}}$ and $\Delta_0 (d) = \left[ A_s \sqrt{\frac{2\pi d d_s}{a^2}} e^{-\frac{d}{d_s}}  - \frac{3\pi C_6 }{8 d^5 a} \right] - \frac{1}{2} U_0 (d)$. Here $d_s$ and $A_s$ are the decay length and magnitude of the short-range interaction, while $C_6$ is the Hamaker constant following the Born-Mayer-Buckingham interatomic potential $A_s e^{-r/d_s} - C_6 r^{-6}$ \cite{Stone2013} (see section S-I in the Supplementary Information \cite{supp} for details). More general, many-body and nonlinear effects can also be taken into account in the particle-chain interaction potential \cite{Le2024, Dang2025CPC}. For a fixed distance $d$, the quantity $\Delta_0 (d)$ is simply a constant and can be dropped out from the above Hamiltonian since it only results in a global shift of the energy spectrum with no other consequences to the physics of the problem. The center of the driving optical trap is $x_c = vt$.

The above Hamiltonian can be re-written in a more convenient form,
\begin{equation}
    \label{eqn:HQ-dimensionless}
    \hat{H}_S (t) = - \frac{1}{2} \frac{\partial^2}{ \partial \overline{x}^2} + \frac{1}{2} \left( \overline{x} - \overline{v} \overline{t} \right)^2 + u_0 \sin^2 \left( \frac{\pi \ell}{a} \overline{x} \right) ,
\end{equation}
where $u_0 = \frac{U_0(d)}{\hbar \Omega}$, $\overline{x} = \frac{x}{\ell}$, $\overline{t} = \frac{t}{\tau}$ and $\overline{v} = \frac{v}{\nu}$ are unitless quantities expressed in terms of a characteristic length $\ell = \sqrt{\frac{\hbar}{M\Omega}}$, period of motion $\tau = \frac{1}{\Omega}$, and characteristic velocity $\nu = \frac{\ell}{\tau} = \sqrt{\frac{\hbar\Omega}{M}}$. This rescaling is equivalent to setting $\hbar = M = \Omega = 1$.     

The above Hamiltonian is a quantum mechanical version of the classical Prandtl-Tomlinson model of a particle driven by a spring above a sinusoidally corrugated potential. This model has found many applications in the analysis of dynamic friction, especially in the understanding of experiments using atomic force microscopes \cite{Meyer1998}. It has also been utilized in capturing temperature in sliding frictional processes \cite{Muser2011}. Nevertheless, this is a classical approach, and here we expand its applicability at the quantum mechanical level. 

\subsection{Dynamics of the moving quantum particle: unitary evolution}
To investigate the dynamical evolution of this closed system, we first solve the eigenvalue problem of the Schr\"odinger equation,
\begin{eqnarray}
    \label{eqn:quasistatic}
    \hat{H}_S (\overline{t}) \left| n (\overline{t}) \right> = E_n (\overline{t}) \left| n (\overline{t}) \right>. 
\end{eqnarray}
For this purpose, the eigenstates $ \left|n (\overline{t}) \right>$ are represented in terms of the eigenfunctions of a one-dimensional harmonic oscillator with a time-dependent displacement $\left| n^{(0)} (\overline{t}) \right>  = \frac{e^{-(\overline{x} -\overline{v} \overline{t} )^2/2}}{\sqrt{2^n n! \sqrt{\pi}}}  H_n \left( \overline{x} - \overline{v} \overline{t} \right)$ where $H_p$ are $p$-th Hermite polynomials \cite{Sakurai2020}. The  eigenstates  are then decomposed as $\left| n (\overline{t}) \right> = \sum\limits_{p=0}^{\infty} c_{n,p} (\overline{t}) \left| p^{(0)} (\overline{t}) \right>$ with $\sum\limits_{p=0}^{\infty}  \left| c_{n,p} (\overline{t}) \right|^2 = 1$. In this representation, Eq. \eqref{eqn:quasistatic} becomes the following eigenvalue equation in a matrix form,
\begin{eqnarray}
    \label{eqn:eigenvalue}
    \sum\limits_{p'=0}^{\infty} (H_S)_{p,p'}( \overline{t}) c_{n,p'} (\overline{t}) =  E_n (\overline{t}) c_{n,p} ( \overline{t}).
\end{eqnarray}
The Hamiltonian matrix elements $(H_S)_{n,n'}$ are found as
\begin{eqnarray}
    \label{eqn:HQ-matrix}
    (H_S)_{n,n'}(\overline{t}) = \left< n^{(0)} (\overline{t}) \right| \hat{H}_S (t) \left| n'^{(0)} (\overline{t}) \right> = \left(n + \frac{1}{2} + \frac{1}{2} u_0 \right) \delta_{n,n'} + u_0 \mathcal{V}_{n,n'} (\overline{t}) .
\end{eqnarray}
\begin{eqnarray}
    \label{eqn:Vcos-method-2}
     \mathcal{V}_{n,n'} (\overline{t}) =  \frac{(-1)^{\left\lfloor\frac{|n-n'|}{2} \right\rfloor}}{\sqrt{2^{|n-n'|+2}}} &&  \sqrt{\frac{\min(n,n')!}{\max(n,n')!}} \left( \frac{2\pi \ell}{a} \right)^{|n-n'|} e^{-\frac{1}{4} (\frac{2\pi \ell}{a})^2} \mathcal{L}_{\min(n,n')}^{|n-n'|} \left[ \frac{1}{2} \left(\frac{2\pi \ell}{a} \right)^2 \right] \nonumber\\
    && \times \left\{ \begin{array}{ll}
    + \sin\left( \frac{2\pi \ell}{a} \overline{v} \overline{t}\right) & \text{when } |n-n'| \text{ is odd} \\
    - \cos\left( \frac{2\pi \ell}{a} \overline{v} \overline{t} \right) & \text{when } |n-n'| \text{ is even} 
    \end{array} \right.,
\end{eqnarray}
where $\mathcal{L}_{a}^b (x)$ are the associated Laguerre polynomials. Note that in this basis representation, the Hamiltonian matrix is given in an entirely analytical form. Its diagonal elements contain the time-independent eigenenergies of a  harmonic oscillator shifted by a constant value.  The time evolution is captured in the matrix element $ \mathcal{V}_{n,n'} (\overline{t})$ containing a decaying oscillatory-like functions with period $a/\ell$. See in section S-II in the Supplementary Information \cite{supp} for derivation of these matrix elements.
      
Eqs.\eqref{eqn:eigenvalue}-\eqref{eqn:Vcos-method-2} can now be solved numerically to find the time-dependent eigenvalues $ E_n ( \overline{t})$ of the Hamiltonian of the system with corresponding eigenvectors labeled as $\left[ c_{n,0} ( \overline{t}), c_{n,1} (\overline{t}), c_{n,2} ( \overline{t}), \ldots \right]^{T}$. The infinite dimension of the Hamiltonian matrix is problematic for the numerical calculations, thus restrictions to a large enough matrix are imposed. As we focus on the lowest five eigenenergies, the eigenvalue problem is solved for an $H_S$ matrix with 25 elements, which is large enough for convergence. More details about the diagonalization of $\hat{H}_S$ and the numerical procedure can be found in section S-III in the Supplementary Information \cite{supp}.

The dynamical evolution of this closed system can be described by the density matrix operator $\hat{\rho}_S (\overline{t})$  governed by the Liouville-von Neumann equation \cite{Yamaguchi2017}. Utilizing the basis set of the time-dependent harmonic oscillator $ \left| n^{(0)} (\overline{t}) \right>$, this equation can be cast into 
\begin{equation}
    \label{eqn:rho-ODE}
    \dfrac{d}{d \overline{t}} \rho_S (\overline{t}) = -i \left[ H_S (\overline{t}) - \sigma_t (\overline{t}) , \rho_S (\overline{t}) \right],
\end{equation}
with matrix elements $\left( \rho_S \right)_{n,n'} (\overline{t}) =  \left< n^{(0)} (\overline{t}) \right| \hat{\rho}_S (\overline{t}) \left| n'^{(0)} (\overline{t}) \right>$ and $\left( \sigma_t \right)_{n,n'} (\overline{t}) =  i \left< n^{(0)} (\overline{t}) \right| \dfrac{d}{d \overline{t}} \left| n'^{(0)} (\overline{t}) \right> $. The explicit expression of the Hermitian matrix $\left( \sigma_t \right)_{n,n'} (\overline{t})$ is given in section S-IV in the Supplementary Information \cite{supp}. {Here, the matrix representation of any operator (denoted with a hat) in the basis set of the time-dependent harmonic oscillator does not carry a hat.} To solve the above equation, we utilize {the fourth-order} Runge-Kutta method \footnote{The considered time domain for the moving particle (from $t_{min} = 0$ to $t_{max} = 3T$ in this study) is divided evenly by infinitesimally small intervals $\Delta t$ {which} must be smaller than the smallest period of Bohr oscillation between eigenstates $\Delta t < \left|\text{max}(E_n (\overline{t}) - E_m (\overline{t})) \right|^{-1}$}. 

The evolution of this system is studied by tracking several properties defined with the  density matrix operator. These include the instantaneous average energy of the system $\left< E \right> (\overline{t})$, the eigenstate population $P_n(\overline{t})$, and linear entropy $S_L (\overline{t})$ conveniently defined in  Table \ref{tab:1}. As $\left< E \right>$ changes in time, $P_n(\overline{t})$ shows the different eigenstate contributions in the dynamical state of the moving particle at each point in time. Also, $S_L (\overline{t})$ measures the states mixture as a function of time \cite{Peters2004}. We can also track the time-dependent global geometric phase of the system determined as \cite{Tong2004}, 
\begin{eqnarray}
\label{eqn:phase-function}
    & \gamma (\overline{t}) = \textrm{arg} \left[ \sum_k \sqrt{  \xi_k (0)  \xi_k (\overline{t}) } \left< \phi _k (0) \right. \left| \phi _k (\overline{t}) \right> \exp{\left( - \int_0^{\overline{t}}  \left< \phi _k (0) \middle| \frac{d  \phi_k}{d \overline{t}} (\overline{t}) \right> d \overline{t}  \right)} \right]. &
\end{eqnarray}
where {$\xi_k(\bar{t})$} and $\left| \phi _k (\overline{t}) \right>$ are instantaneous eigenvalues and associated eigenvectors of the density matrix operator, such that $\hat{\rho}_S = \sum_k \xi_k(\bar{t})) \left| \phi _k (\overline{t}) \right> \left< \phi _k (\overline{t}) \right|$. The quantity $\gamma (\overline{t})$ is exclusively associated with the geometric path of the quantum mechanical state in time.

\begin{table}[htbp]
    \centering
    \caption{Quantum mechanical properties of the coupled moving particle/atomic chain system. Here the linear entropy is renormalized by a factor $\frac{N_{max}}{N_{max}-1}$ where $N_{max}$ is the cutoff size of the Hamiltonian as suggested in Ref.  \cite{Peters2004}. The standard deviation of an arbitrary observable $\hat{a}$ is $\sigma_a = \sqrt{\text{Tr} \left[ \hat{a}^2 \hat{\rho}_S (\overline{t}) \right] - \text{Tr} \left[ \hat{a} \hat{\rho}_S (\overline{t}) \right]^2}$. The work associated with the moving harmonic trap $\mathcal{W}(\overline{t})$ is calculated from the following integral $\mathcal{W}(\overline{t}) = \int_0^{\overline{t}} \text{Tr} \left[ \hat{\rho}_S (\overline{t}^{\prime}) \frac{d \hat{H}_S (\overline{t}^{\prime}) }{d \overline{t}^{\prime}} \right] d \overline{t}^{\prime} $ . }
    \label{tab:1}
    \begin{ruledtabular}
    	\resizebox{\textwidth}{!}{
    \begin{tabular}{lll}
    Internal properties  & Kinematic properties  & Dynamic properties  \\
    \hline & & \\
    Average energy $\left< E \right> (\overline{t}) = \text{Tr} \left[ \hat{\rho}_S (\overline{t}) \hat{H}_S (\overline{t}) \right]$ 
     & Average position $\left< x \right> (\overline{t}) = \text{Tr} \left[ \hat{x} \hat{\rho}_S (\overline{t}) \right] $ &  Heat transfer $Q = \Delta \left< E \right> (\overline{t}) - \mathcal{W} (\overline{t})$  \\
   Eigenstate population $P_n(\bar{t})=\left< n(\bar{t}) \right| \hat{\rho}_S(\bar{t}) \left| n(\bar{t}) \right>$
         & Average velocity $\left<v\right> (\overline{t}) = M^{-1} \text{Tr} \left[ \hat{p} \hat{\rho}_S (\overline{t}) \right]$ &  Power transfer $P = Q/\overline{t}$  \\
   Linear entropy $S_L(\overline{t}) = \frac{N_{max}}{N_{max}-1}\left( 1 - \text{Tr} \left( \hat{\rho}_S^2 \right) \right)$
       & Heisenberg Principle $\sigma_x (\overline{t}) \sigma_p (\overline{t})$  
       &  Lateral force $\left<F_L\right> (\overline{t}) = -M \Omega^2 \left(\left<\overline{x}\right>(\overline{t}) - \overline{v} \overline{t} \right)$  \\
    Geometric phase $\gamma (\overline{t})$ as in Eq. \eqref{eqn:phase-function}  & 
    \end{tabular}
    }
    \end{ruledtabular}
\end{table}

Before presenting results from the numerical calculations, we examine some limiting cases as dictated by the properties of the Hamiltonian. Eqs. \eqref{eqn:HQ-matrix}, \eqref{eqn:Vcos-method-2} show that the characteristic behavior of the matrix elements $\mathcal{V}_{n,n} (\overline{t})$ is controlled by the parameters $\dfrac{2\pi\ell}{a}$ and the velocity $\overline{v} = \dfrac{v}{\nu}$. In the case of 
$a \ll \ell$, the asymptotic behavior of the exponential term  $e^{-\frac{1}{4} (\frac{2\pi \ell}{a})^2} \to 0$ makes the last term in the Hamiltonian matrix \eqref{eqn:HQ-matrix} negligible. Consequently,  the eigenenergies of the system are simply the diagonal elements of $H_S$: $E_n (\overline{t}) \approx n + \dfrac{1}{2} + \dfrac{u_0}{2}$. This is understood by realizing that the {trapped particle} ``sees" an almost continuous chain due to the very small $a$, thus the particle effectively behaves as a shifted harmonic trap with constant eigenenergies. 

In the case of $a \gg \ell$, the $\mathcal{V}_{n,n} (\overline{t})$ off-diagonal elements diminish since $\left( \dfrac{2\pi \ell}{a} \right)^{|n-n'|} \to 0$ and only the diagonal terms matter. As a result, the eigenenergies become oscillatory according to $E_n (\overline{t}) \approx n + \dfrac{1}{2} + \dfrac{u_0}{2} - \dfrac{u_0}{2} \cos \left( \frac{2 \pi \ell}{a} \overline{v} \overline{t} \right)$. This situation corresponds to the harmonic trap of the particle being much narrower than the periodic potential of the atomic chain. The particle is effectively trapped in the minimum of the potential as it moves along the chain.  In both limits, $\left( \dfrac{2\pi \ell}{a} \right) \to 0$ or $\left( \dfrac{2\pi \ell}{a} \right) \to \infty$, the gaps between the eigenenergies are time-independent. 

When the ratio $\left( \dfrac{2\pi \ell}{a} \right) \sim 1$, however, the off-diagonal elements play a vital role for the particle as the eigenstates now experience the influence of different $\left| n^{(0)} (\overline{t}) \right> $. For example, it is possible that the potential of the system evolves from a harmonic trap with a single minimum to one with a double minima. This occurs when the depth of the periodic potential satisfies $u_0 > 2 \left( \dfrac{a}{2 \pi \ell} \right)^2$. As a result, there are anti-crossings between eigenenergy levels signaling the occurrence of Landau-Zener tunneling \cite{Landau1932, Zener1932} whose probability is also controlled by the velocity $\overline{v}$. One notes that the condition supporting Landau-Zener tunneling $u_0 > 2 \left( \dfrac{a}{2 \pi \ell} \right)^2$ is consistent with the condition for the occurrence of stick-slip motion in the classical Prandtl-Tomlinson model \cite{Socoliuc2004, Gangloff2015, Counts2017} defined through the  corrugation parameter $\eta = \frac{u_0}{2} \left( \frac{2\pi \ell}{a} \right)^2 > 1 $. Unlike the classical Prandtl-Tomlinson model, multiple stick-slip motion requires constraints not only through the  corrugation parameter $\eta$ but also the ratio $\left( \dfrac{2\pi \ell}{a} \right)$. With a fixed corrugation parameter $\eta$, the shape of the potential surface of the particle at $t = T/2$ is invariant but the ratio $\left( \dfrac{2\pi \ell}{a} \right)$ controls the scale of that potential surface. As a known result from quantum mechanics for the double-well potential, the depth of the wells would tell us how many energy levels stay inside the wells and the numbers of the anti-crossing levels (see Ref. \cite{Le2018} for example). If there are no anti-crossing levels (when $\left( \dfrac{2\pi \ell}{a} \right)$ is too large or too small), there is no Landau-Zener diabatic transition and consequently there is no stick-slip motion.

\begin{figure}[htpb]
	\centering
	\includegraphics[width = 1.0 \textwidth]{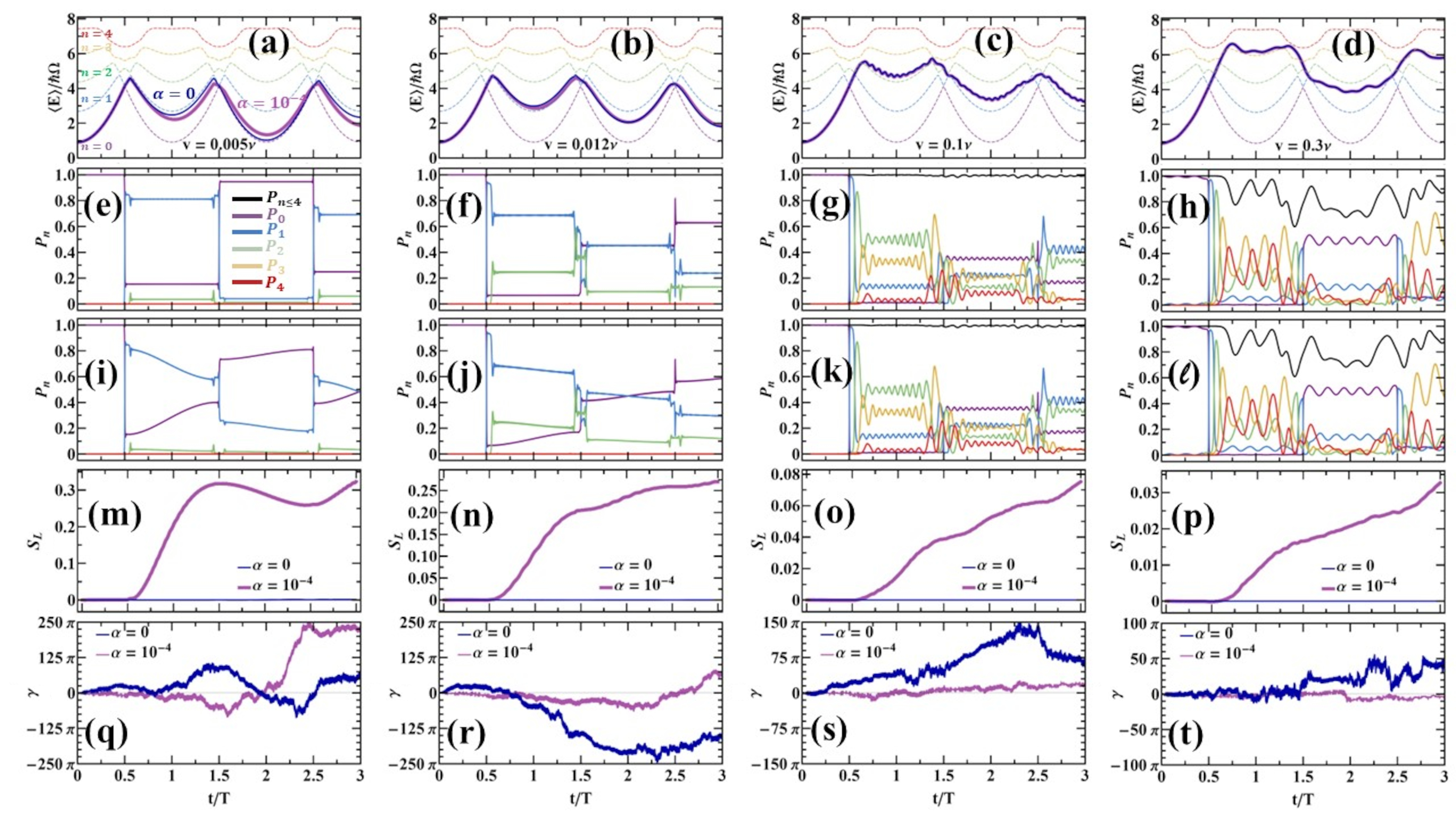}
	\caption{\label{fig:Figure2} 
		The average energy $\left< E \right>$ and the first five eigenergies of Hamiltonian $\hat{H}_S(t)$ normalized by $\hbar\Omega$ as a function of time rescaled by the period $T= a /v$ for (a) $v = 0.005 \nu$, (b) $v = 0.012 \nu$, (c) $v = 0.1 \nu$ and (d) $v = 0.3 \nu$. The population of the first five eigenstates of $\hat{H}_S(t)$ as a function of time $t/T$  for (e) $v = 0.005 \nu$, (f) $v = 0.012 \nu$, (g) $v = 0.1 \nu$ and (h) $v = 0.3 \nu$. The population of the first five eigenstates of $\hat{H}(t)$ as a function of time $t/T$ for (i) $v = 0.005 \nu$, (j) $v = 0.012 \nu$, (k) $v = 0.1 \nu$ and (k) $v = 0.3 \nu$. The linear entropy as a function of time $t/T$ for $\hat{H}_S(t)$ and $\hat{H}(t)$ for (m) $v = 0.005 \nu$, (n) $v = 0.012 \nu$, (o) $v = 0.1 \nu$ and (p) $v = 0.3 \nu$. The geometric phase as a function of time $t/T$ for $\hat{H}_S(t)$ and $\hat{H}(t)$ for (q) $v = 0.005 \nu$, (r) $v = 0.012 \nu$, (s) $v = 0.1 \nu$ and (t) $v = 0.3 \nu$. Here the distance between the particle and the chain is $a = 2 \pi \ell$, the depth of the periodic potential is $U_0 = 5 \hbar \Omega$. The temperature and cutoff energy of the harmonic bath are $k_B T = 0.01 \hbar \Omega$, $\hbar \omega_c = 50 \hbar \Omega$, and the particle-bath coupling constant in $\bar{H}(t)$ is taken as $\alpha = 10^{-4}$. }
\end{figure}

In Fig. \ref{fig:Figure2}(a-d), we show the evolution of the lowest five eigenenergies of the particle for different velocities together with their population dynamics in Fig. \ref{fig:Figure2}(e-h) by focusing on the Landau-Zener tunneling regime. For this purpose, we take $\dfrac{a}{2\pi \ell} = 1$ and $u_0 = 5$, which corresponds to a {classical corrugation parameter} $\eta = 2.5 > 1$. From the theory of Landau-Zener tunneling \cite{Landau1932, Zener1932}, the probability of a diabatic transition at the anti-crossing between levels $n$ and $n^{\prime}$ is estimated as $P_{n \to n'} = \exp{\left( - \dfrac{v_{n,n'}}{v} \right)}$ where $v_{n,n'} = \frac{\pi \left|E_n - E_{n'}\right|^2}{2 \frac{\partial |E_n - E_{n'}|}{\partial x_c}}$ 
with $x_c = vt$ being the position of the center of the optical trap. Thus for $\dfrac{a}{2\pi \ell} = 1$ and $u_0 = 5$, we estimate that $v_{01} = 8.39 \times 10^{-4} \nu$, $v_{12} = 1.75 \times 10^{-2} \nu$, $v_{23} = 1.25 \times 10^{-1} \nu$ and $v_{34} = 4.50 \times 10^{-1} \nu$.

By tracking  $\left< E \right> (\overline{t})$ in Fig. \ref{fig:Figure2}a for $v = 0.005 \nu$, we find that the particle starts out in the ground state of $\hat{H}_S$. At $t/T = 0.5$  the anti-crossing between $n=0$ and $n=1$ states induces a diabatic tunneling transition with probability of $P_{0 \to 1} \approx 85 \%$ (Fig. \ref{fig:Figure2}e). Afterwards, the particle is in a superposition composed of the $n=0$ and $n=1$ states since higher-level transitions are negligible. The two states are disentangled by a diabatic transition from $n = 1$ to $n = 0$ at $t/T = 1.5$. Indeed, Fig. \ref{fig:Figure2}e shows that for $t/T = (0.5, 1.5)$ the level population contains about 82 $\%$  from the first excited state and about 15 $\%$ from the ground state, while the remaining part belongs to the second excited state. In the interval $t/T = (1.5, 2.5)$, the population is composed of about 95 $\%$ contributions from the ground state and the diabatic transition at $t/T = 2.5$ redistributes the populations with about 23 $\%$, 70  $\%$ and 7 $\%$ for $n = 0,1,2$ states. 

Increasing the velocity to $v = 0.012 \nu$ activates tunneling between $n=0 \to n=1$ and $n=1 \to n=2$ at $t/T \sim 0.5$ and between $n=2 \to n=3$ at $t/T \approx 0.6$ with corresponding  transition probabilities $P_{0 \to 1} = 93 \%$, $P_{1 \to 2} = 23 \%$ and $P_{2 \to 3} = 0.003 \%$.  As a result, there are various redistributions in the level populations, as shown in Fig. \ref{fig:Figure2}f. At $v=0.1\nu$ and $v=0.3\nu$ velocities, tunneling involving higher energetic states with increasing probabilities are further activated. Thus,  the particle experiences the {Landau-Zener} tunneling effects from {higher-lying} energy levels ((Fig. \ref{fig:Figure2}c,g) and Fig. \ref{fig:Figure2}d,h).  This slow-to-fast moving particle progression shown in Fig. \ref{fig:Figure2} is accompanied by transitioning to more complicated time-dependent tunneling between higher states with complex patterns of periodicity.

\subsection{Dynamics of the moving quantum particle with coupling to the environment}

The dynamics of the system described by the Hamiltonian in Eqs. \eqref{eqn:HQ}-\eqref{eqn:HQ-dimensionless} corresponds to a coherent evolution of a closed system and it exemplifies internal quantum mechanical effects in the motion. The forced motion of the particle, however, results in energy removed by dissipation. To model this we consider that the {particle}/atomic chain system is coupled to an external thermostat taken as a harmonic bath in the Caldeira–Leggett model \cite{Caldeira1981, Weiss2012, Zanca2018} 
\begin{eqnarray}
    \label{eqn:H-full}
    \hat{H}(t) = \hat{H}_S (t) + \hat{H}_{int+B}, \quad     \hat{H}_{int+B} =  \sum_i \left[ -\frac{1}{2 m_i} \frac{\partial^2}{ \partial x_i ^2}  + \frac{1}{2} m_i \omega_i^2 \left( x_i - \frac{c_i}{m_i \omega_i^2} \sin \left( \frac{2\pi \ell}{a} \overline{x} \right) \right) ^2 \right],
\end{eqnarray}
where $x_i$ are the spatial degrees of freedom of the bath oscillation modes  with  parameters $m_i, \omega_i, c_i$ for their mass, frequency, and strength, respectively. The bath is characterized by a spectral function  $J(\omega) = \sum\limits_i \frac{c_i^2}{2 m_i \omega_i} \delta \left( \omega - \omega_i \right)$ which describes the dissipative coupling with the system  \cite{Dittrich1998, Weiss2012}. Here we focus on the Ohmic dissipation limit with $J(\omega) = 2 \alpha \omega e^{-|\omega| / \omega_c}$, where $\omega_c$ is a cutoff frequency separating the low-frequency regime for which $J(\omega) \propto \omega$ and the high-frequency regime specified by a decaying Ohmic damping. The dimensionless constant $\alpha \ll 1$ specifies the coupling between the particle/chain system and the harmonic bath. Notice that if $\alpha = 0$, then $J(\omega) = 0$ for all frequencies and $c_i = 0$ i.e. there is no coupling between the system and the bath and $\hat{H}_{int} = 0$. 

To continue further, the Hamiltonian in Eq. \eqref{eqn:H-full} is represented by separating the degrees of freedom of the  nanoparticle from the degrees of freedom of the harmonic bath, such that
\begin{eqnarray}
   \label{eqn:Comp-Hamilt}
     \hat{H}(t) = \hat{H}_S^{\prime} (\overline{t}) \otimes \hat{1}_B + \hat{A} \otimes \hat{B} + \hat{1}_S \otimes \hat{H}_B,
\end{eqnarray}
with a renormalization operator $\hat{A} = \sin \left( \frac{2\pi \ell }{a} \overline{x} \right)$ and $\hat{H}_S^{\prime} (\overline{t}) = \hat{H}_S (\overline{t}) +  u_{rn} \hat{A}^2$ act in the subspace of the particle, while $\hat{B} = - \sum\limits_i c_i x_i $ and $\hat{H}_B = \sum\limits_i \left( -\frac{1}{2 m_i} \frac{\partial^2}{ \partial x_i ^2}  + \frac{1}{2} m_i \omega_i^2 x_i^2 \right)$ operate in the subspace of the harmonic bath. The term $ u_{rn} \hat{A}^2$ captures the dissipative coupling between the moving particle and the heat bath.  The renormalization constant can be determined via the spectral density function $u_{rn} = \sum\limits_i \frac{c_i^2}{2m_i \omega_i^2} = \sum\limits_i \frac{c_i^2}{2m_i \omega_i} \int_{0}^{+\infty} \omega_i^{-1} \delta (\omega - \omega_i) d \omega   = \int_0^{+\infty} \omega^{-1} J(\omega) d \omega = 2 \alpha \omega_c$, as also shown in Ref.\cite{Zanca2018}. This renormalization potential in the total Hamiltonian \eqref{eqn:H-full} ensures that the system maintains its global minimum regardless of the bath \cite{Weiss2012}.

Assuming weak coupling consistent with the Born-Markov approximation, the quantum Liouville - von Neumann master equation for the density matrix is written as \cite{Yamaguchi2017}
\begin{eqnarray}
    \label{eqn:QLV-WCA}
\frac{d}{d \overline{t}} \hat{\rho }_S (\overline{t})  = - i \left[ \hat{H}_S (\overline{t}) + 2 \alpha \omega_c \hat{A}^2, \hat{\rho}_S (\overline{t}) \right] - \left\{ \left[ \hat{A}, \hat{S} (\overline{t}) \hat{\rho}_S (\overline{t}) \right] + \text{h.c.} \right\},
\end{eqnarray}
where the bath-convoluted operator $\hat{S} (t)$ is defined as $\hat{S} (t) \approx \sum\limits_{m,n} \left< n(\overline{t}) \middle| \hat{A} \middle| m(\overline{t}) \right> \Gamma \left( E_{m}-E_{n}\right) \allowbreak \left|n(\overline{t}) \right> \left< m(\overline{t}) \right|$ in which the bath-induced transition rate $\Gamma \left( E \right)$ is simply the Fourier transform of the bath correlation function $C (\tau)$ for which $\Gamma (E) = \int_{0}^{+\infty} d\tau C (\tau) e^{ i (E +i 0^+)\tau}$. Taking into account that the harmonic bath is in thermal equilibrium obeying the Bose-Einstein distribution $f_{BE} (\omega) = \left(e^{\omega/k_B T} - 1 \right)^{-1}$, the bath correlation function can be expressed {in terms of} the spectral function $C (\tau) =\int_{-\infty}^{+\infty} e^{i \omega \tau} f_{BE} (\omega) J(\omega) d \omega$.  The bath-induced transition rate can further be simplified as $\Gamma \left( E \right) = \frac{1}{2} \gamma (E) + i \sigma (E)$ where $\gamma (E) = \frac{4 \pi \alpha E e^{-|E|/\omega_c}}{1 - e^{-E/{k_B T}}}$ and $\sigma (E) = \mathcal{P} \int_{-\infty}^{+\infty} \frac{ 2 \alpha \xi e^{-|\xi|/\omega_c} d\xi}{(E + \xi) (e^{-\xi/k_B T} - 1 )} $. For $E = 0$, $\gamma (0) = 4\pi \alpha k_B T$ and $\sigma (0) = - 2 \alpha \omega_c$. 

Similar to the case of the moving particle with no external bath, Eq. \eqref{eqn:QLV-WCA} can be expressed in the basis set of the time-dependent harmonic oscillators $ \left| n^{(0)} (\overline{t}) \right> = \varphi_n^{HO}(\bar{x}-\bar{v}\bar{t})$:
\begin{eqnarray}
    \label{eqn:QLV-WCA-ODE}
\frac{d}{d\overline{t}} \rho_S (\overline{t})  = -i \left[ H_S (t) + 2 \alpha \omega_c A^2 (\overline{t}) - \sigma_t (\overline{t}),  \rho_S (\overline{t})  \right] - \left\{ \left[ A (\overline{t}), S (\overline{t}) \rho_S (\overline{t}) \right] + \text{h.c.}  \right\},
\end{eqnarray}
where the matrix elements of $\hat{A}$ and $\hat{S} (\overline{t})$ are given as follows
\begin{eqnarray}\label{eqn:A-oper-BM-1D}
     A_{n,n'} (\overline{t}) && = \left< n^{(0)}(\overline{t}) \middle|  \sin \left( \frac{2\pi \ell}{a} \overline{x} \right) \middle| n'^{(0)}(\overline{t}) \right>  \nonumber\\
     && =  \frac{ (-1)^{ \left\lfloor \frac{|n-n'|}{2} \right\rfloor } }{\sqrt{2^{|n-n'|+2}}} \sqrt{\frac{\min(n,n')!}{\max(n,n')!}} \left( \frac{2\pi \ell}{a} \right)^{|n-n'|} e^{-\frac{1}{4} (\frac{2\pi \ell}{a})^2} \mathcal{L}_{\min(n,n')}^{|n-n'|} \left[ \frac{1}{2} \left(\frac{2\pi \ell}{a} \right)^2 \right] \nonumber\\
    &&  \quad \quad \quad \quad \quad \quad \times \left\{ \begin{array}{ll}
    \cos \left( \frac{2\pi \ell}{a} \overline{v} \overline{t}\right) & \text{when } |n-n'| \text{ is odd} \\
    \sin \left( \frac{2\pi \ell}{a} \overline{v} \overline{t} \right) & \text{when } |n-n'| \text{ is even} 
    \end{array} \right.,\\
    \label{eqn:S-oper-BM-1D}
    S_{n,n'} (\overline{t}) && =  \sum_{m,m',k, k'} c_{m,n} (\overline{t}) c^{*}_{m,k} (\overline{t}) A_{k,k'}(\overline{t}) c_{m',k'}(\overline{t}) c_{m',n'}^{*} (\overline{t}) \Gamma\left[ E_{m'}(\overline{t})-E_{m}(\overline{t})\right]  .
\end{eqnarray}
Details about the derivation of $\hat{H}(t)$ and the quantum Liouville-von Neumann equation are given in section S-V in the Supplementary Information \cite{supp}.

The dynamic evolution of the particle is now examined based on Eqs. \eqref{eqn:Comp-Hamilt},\eqref{eqn:QLV-WCA} by turning on the coupling to the bath with parameters $\alpha = 10^{-4}$, $\omega_c = 50 \Omega$ and $k_B T = 0.01 \hbar \Omega$. Here, we choose $\alpha$ small enough to maintain the validity of the Born-Markov approximation, while $\omega_c$ is large enough to include higher energy level transitions in the system. {We note that the positivity of the density matrix $\hat{\rho}_S$ is preserved following the arguments in \cite{Yamaguchi2017}. Since the condition $\min\limits_{\epsilon \neq \epsilon^{\prime}} \left| \epsilon - \epsilon^{\prime} \right| \gg \max \gamma (\epsilon)$, where $\epsilon, \epsilon^{\prime}$ are possible values of the differences between the instantaneous energy levels $E_{n}(t), E_{m}(t)$, is fulfilled for the solution of \eqref{eqn:QLV-WCA}, the non-Lindblad terms can be neglected under the rotating wave approximation, and the quantum master equation  is  positive (see details in the Supplementary Information \cite{supp}). In our numerical calculations, the two smallest Bohr frequencies are $\epsilon = \left| E_{1}(t) - E_0(t) \right| \approx 0.045 \hbar \Omega $ and $\epsilon'= \left| E_{2}(t) - E_1(t) \right| \approx 0.178 \hbar \Omega $; hence, $\min\limits_{\epsilon \neq \epsilon^{\prime}} \left| \epsilon - \epsilon^{\prime}\right| \approx 0.178 \hbar \Omega - 0.045 \hbar \Omega = 0.133 \hbar \Omega$. On the other hand, $\max \gamma (\epsilon) \approx 4 \pi \alpha \hbar \omega_c = 0.063 \hbar \Omega < \min\limits_{\epsilon \neq \epsilon^{\prime}} \left| \epsilon - \epsilon^{\prime} \right|$. Therefore, the positivity condition holds for this choice of parameters. }


In Fig.\ref{fig:Figure2} (a-d), we show the numerical calculations for the ground state average energy for the system/bath Hamiltonian. The $\alpha=0$ and $\alpha=10^{-4}$ results are very similar, especially for larger velocities. The role of the bath can be better discerned by looking at the level population dynamics. Fig. \ref{fig:Figure2}(i-l) shows that for $t/T < 0.5$ the particle remains in its ground state for practically all studied velocities. At 
$t/T \approx 0.5$ the particle experiences tunneling and it acquires higher level energy states. For the $v = 0.005 \nu$ case, for example, Fig. \ref{fig:Figure2}(i) shows that in the  $t/T=(0.5, 1.5)$ interval, the ground and excited state admixture changes in time in such a way that both contributions become of similar value at the end of the interval - $40\%$ from the ground state and $60\%$ for the first excited state. This type of admixture dynamics becomes more complex for higher velocities. 

Clearly, the disorder in the system has increased after the initial tunneling at $t/T\approx 0.5$ due to the coupling with the external bath. This correlates well with the linear entropy in Fig.\ref{fig:Figure2} (m-p), which shows that $S_L$ increases over time when the system is coupled to the external thermostat, while $S_L$ remains zero for the closed system.

Signatures of the {Landau-Zener} tunneling and coupling to the environment can also be found in the geometric phase, shown in Fig. \ref{fig:Figure2}(q-t). The different crossings involved in the dynamics of the system can cause a complex behavior of  $\gamma(\bar{t})$ leading to interference of transition path interference due to different phase accumulations \cite{Shevchenko2010}. 
Experimental interferometry measurements can access the {Landau-Zener} tunneling geometric phase in different atomic and optical systems  \cite{Bouwmeester96,Gasparinetti2011}. Fig. \ref{fig:Figure2}(q-t) shows that for the unitary evolution of the system, $\gamma(\bar{t})$ is strongly dependent on the velocity of the particle. The geometric phase experiences different behavior in time and it can even change sign. Turning on the coupling to the bath adds dissipation which further adds to the complexity of $\gamma$. Nonunitary effects have been studied in various two-level systems \cite{Viotti2022,Lombardo2006,Viotti2023} {or in the context of Casimir friction \cite{Decca2020, Lombardo2021}}, demonstrating that $\gamma$ has a stochastic character due to occurrence of random quantum phase jumps. Fig. \ref{fig:Figure2}(q-t) further shows that the frictional dissipation has a significant influence on the geometric phase of the particle for each studied velocity adding to the complex behavior of $\gamma(\bar{t})$.

\subsection{Dynamics of the moving particle within the classical Prandtl-Tomlinson model}

In order to understand better the quantum mechanical effects in the frictional dissipation process, we also examine the motion of the particle within the classical Prandtl-Tomlinson model. From the classical version of the Hamiltonian in \eqref{eqn:H-full}, the stochastic dynamics of the particle is captured through the Hamilton's equations of motion with coupling to an external bath via a function $f_{bath}$ \cite{Weiss2012}
\begin{eqnarray}
    \label{eqn:hamil-equations}
\left\{    
\begin{array}{lll}
 \dfrac{d^2 \overline{x}}{d\overline{t}^2} + \left( \overline{x} - \overline{v} \overline{t} \right) +\dfrac{  u_0 \pi\ell}{a} \sin \left( \dfrac{2\pi \ell}{a} \overline{x} \right)  &  = & f_{bath} (\overline{x}, \overline{t} ), \\
 \dfrac{d^2 x_i}{d \overline{t}^2} + \omega_i^2 x_i & = & \dfrac{c_i}{m_i} \sin \left( \dfrac{2\pi \ell}{a} \overline{x} \right)  
\end{array}
\right.,
\end{eqnarray}
\begin{eqnarray}
\label{eqn:bath-force}
    f_{bath} (\overline{x}, \overline{t} ) = \dfrac{2\pi \ell}{a} \left( \sum\limits_i x_i c_i \right) \cos \left( \dfrac{2\pi \ell}{a} \overline{x} \right) - \dfrac{2\pi \ell}{a} \left( \sum\limits_i \dfrac{c_i^2}{2 m_i \omega_i^2} \right) \sin \left( \dfrac{4\pi \ell}{a} \overline{x} \right) .
\end{eqnarray}
To solve the above system of equations, 
the bath degrees of freedom $x_i$ are first resolved by using the Green function of the harmonic bath $G(\overline{t}, \overline{t}^{\prime}) = \frac{1}{\omega_i} \Theta (\overline{t} - \overline{t}^{\prime}) \sin\left[ \omega_i (\overline{t} - \overline{t}^{\prime}) \right]$ and taking into account the initial conditions $x_i (0), v_i (0)$ of the external bath. This is followed by a summation of bath degrees of freedom using the spectral function $J(\omega) = \sum\limits_i \frac{c_i^2}{2 m_i \omega_i} \delta \left( \omega - \omega_i \right) = 2 \alpha \omega e^{-|\omega|/ \omega_c}$ and the classical damping correlation $\lim\limits_{\omega_c \to \infty}\frac{4\alpha \omega_c}{1+\omega_c^2 \overline{t}^2} = 4 \pi \alpha \delta (\overline{t}) $ following standard procedure \cite{Weiss2012}. Similar to the Caldeira-Leggett model discussed earlier, $\alpha$ denotes the coupling parameter with the external bath and $\omega_c$ is a cutoff frequency.

The bath coupling function can now be conveniently written as $f_{bath} \left[ \overline{x} (\overline{t}), \overline{t} \right] =f_{vis} \left[ \overline{x} , \dfrac{d \overline{x}}{d\overline{t}} \right] + f_{ran} \left( \overline{t} \right) $, where 
\begin{eqnarray}
\label{eqn:vis-force}
    && f_{vis} \left[ \overline{x} , \dfrac{d \overline{x}}{d\overline{t}} \right] = - \frac{8 \alpha \pi^3 \ell^2}{a^2} \cos ^2\left( \dfrac{2\pi \ell}{a} \overline{x} \right)  \dfrac{d \overline{x}}{d\overline{t}},
     \\
\label{eqn:ran-force}
    && f_{ran} \left(\overline{t} \right) = \dfrac{2\pi \ell}{a} \left\{ \sum\limits_i c_i \left[ x_i^{homo} (\overline{t}) -  \sum\limits_i \dfrac{c_i}{m_i \omega_i^2} \sin \left[ \dfrac{2\pi \ell}{a} \overline{x} (0) \right] \cos\left( \omega_i \overline{t} \right) \right] \right\} \cos \left[ \dfrac{2\pi \ell}{a} \overline{x} (\overline{t}) \right] .
\end{eqnarray}
Since $f_{vis}$ depends on position and velocity of the particle, but not on the initial conditions, this term is interpreted as an environmental viscous force \cite{Weiss2012}. The $f_{ran}$ contribution  depends on the initial conditions of the bath degrees of freedom and denotes the random force due to dissipation from the coupling to the external bath. At thermal equilibrium, the bath follows the canonical Maxwell-Boltzmann distribution, and as a result $f_{ran}$ obeys the fluctuation-dissipation theorem \cite{Rytovbook} such that $\left< f_{ran} (\overline{t}) f_{ran} (\overline{t}^{\prime}) \right>=k_B T \dfrac{8 \alpha \pi^3 \ell^2}{a^3}  \cos^2 \left[ \dfrac{2\pi \ell}{a} \overline{x} (\overline{t}) \right] \delta(\overline{t} - \overline{t}^{\prime})$. Details for the classical Prandtl-Tomlinson model in the presence of an external bath are given in section S-VI in the Supplementary Information \cite{supp}.

\begin{figure}[htbp]
    \centering
    \includegraphics[width = 1.0 \textwidth]{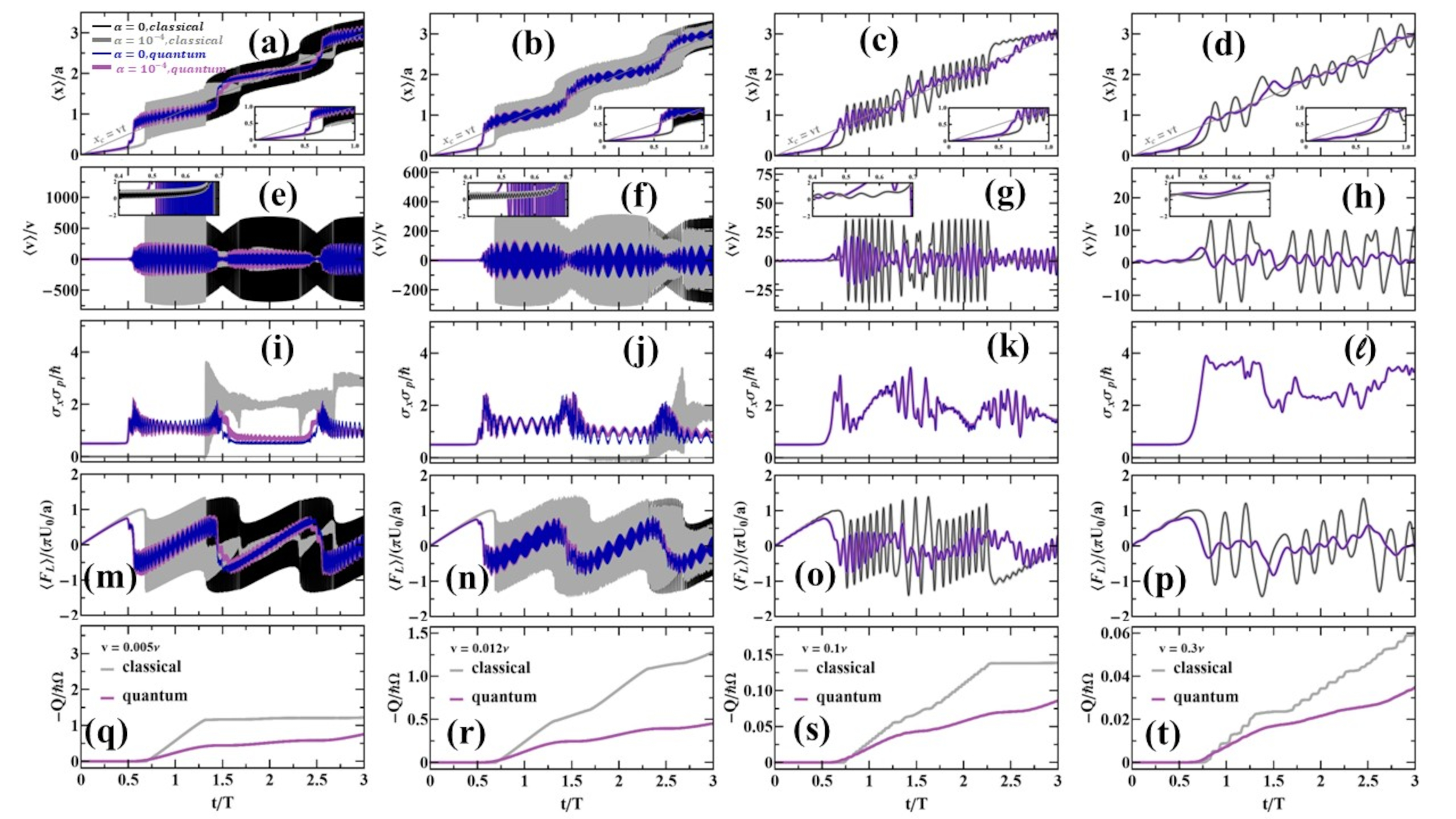}
    \caption{\label{fig:Figure3} Kinematic properties of the moving particle as function of $t/T$. The average displacement $\left<x\right>$ normalized by $a$ for (a) $v=0.005\nu$; (b) $v=0.012\nu$; (c) $v=0.1\nu$; (d) $v=0.3\nu$. The average velocity $\left<v\right>$ normalized by driving velocity of the harmonic trap $v$ for (e) $v=0.005\nu$; (f) $v=0.012\nu$; (g) $v=0.1\nu$; (h) $v=0.3\nu$. Standard deviation product $\sigma_x\sigma_v/\hbar$ for (i) $v=0.005\nu$; (j) $v=0.012\nu$; (k) $v=0.1\nu$; (l) $v=0.3\nu$. Lateral force $\left<F_L\right>$ normalized by $\pi U_0/a$ (m) $v=0.005\nu$; (n) $v=0.012\nu$; (o) $v=0.1\nu$; (p) $v=0.3\nu$. Dissipated heat $-Q$ normalized by $\hbar\Omega$ for (q) $v=0.005\nu$; (r) $v=0.012\nu$; (s) $v=0.1\nu$; (t) $v=0.3\nu$.
    }
\end{figure}

\section{Kinematic and dissipative properties of the moving nanoparticle}

The dissipative evolution of the particle/atomic chain system is further studied by examining its kinematic properties, as given in Table \ref{tab:1}. The moving particle is characterized by its instantaneous average position $\left<x\right> (\overline{t})$ and instantaneous average velocity $\left<v\right> (\overline{t})$. Both are calculated for the closed particle/chain system governed by Eq. \eqref{eqn:HQ} and the particle/chain system with the external bath governed by Eq. \eqref{eqn:H-full}. The results for $\left<x\right> (\overline{t})$ are given in Fig. \ref{fig:Figure3} (a-d). Since the particle is guided forward by the optical trap, its instantaneous position $x_c = v t$ is also shown. The results for $\left<v\right> (\overline{t})$ are given in Fig. \ref{fig:Figure3} (e-h). In order to understand the quantum nature of the motion, the classical position $x_{cl}$ and velocity $v_{cl}$ calculated from the stochastic Newton second law of motion, as previously discussed,  are also calculated and shown in Fig. \ref{fig:Figure3}.

By comparing the different regimes in Fig. \ref{fig:Figure3}(a-d), an emerging common feature is the smooth path in the first half of the period. Analyzing the states of the quantum system shows that in this region the particle remains localized in the minimum of the first well of the overall potential surface regardless of the driving velocity. At $t/T \sim 0.5$ the particle becomes un-stuck and it slips into the second well of the potential surface (see section S-VIII in the  Supplementary Information \cite{supp} for graphical details). In the case of a very slow driving velocity, the particle experiences almost frictionless motion in its initial stick-slip path since it remains in its ground state, as also seen in Fig. \ref{fig:Figure2}(e,i). 

For driving velocities larger than the first Landau-Zener velocity $v \gg v_{01}$, the diabatic tunneling $n=0 \to n=1$ at $t/T\sim 0.5$ makes the state of the particle mainly distribute to the first excited state whose center moves from the second well back to the first well of the potential surface at $t/T = 0.5$. This diabatic Landau-Zener tunneling lengthens the stick-like motion path of the particle, contributing to an increased friction. 

After the initial stick-slip motion, the particle is in a superposition state and it   experiences a series of stick-slip oscillations with a frequency in the order of $\Omega$. Higher driving velocities  involve contributions from higher states, which makes the stick-slip oscillations have more complex oscillatory patterns. Since the period of motion $T$ is shortened for higher $v$, the number of oscillations in one period is reduced. 

While the $\left< x \right>$ and $\left< v \right>$ for the closed and open quantum systems are very similar, there are differences when compared with the classical path. Fig. \ref{fig:Figure3}(a-d) shows that the classical particle also experiences stick-like motion, but at different times. For example, for $v=0.005\nu$, the quantum slip occurs at $t_{slip} \approx 0.49 T$ while the classical slip occurs at $t_{slip} \approx 0.65 T$. This trend also holds for higher driving velocities. A notable difference with the quantum mechanical motion is that the classical particle has to wait until the minimum of the first well of the potential surface completely disappears, which always happens later than $t/T = 0.5$. Since the stick path of the classical motion is much longer than the quantum one, the classical slip starts with higher energy leading to classical stick-slip oscillations with larger amplitudes. This delayed initial stick-slip and larger amplitude oscillations result in an enhanced classical friction compared to the quantum mechanical motion.   
 
Fig. \ref{fig:Figure3}(a-h) further shows that the role of the heat bath is mostly pronounced for smaller driving velocities. Since $T\sim 1/v$, a smaller driving velocity leads to a longer period, hence environmental effects on the particle are better visible.

The instantaneous average velocity of the particle shown in Fig. \ref{fig:Figure3}(e-h) {is consistent} with the displacement behavior. We find that during its initial $t/T<0.5$ time $\left<v\right> \approx 0$. At the slipping moment at $t/T\sim 0.5$, the particle rapidly accelerates catching up with the driving trap. This is followed by oscillatory behavior in $\left< v \right>$ consistent with the oscillations in $\left< x \right>$.  Compared to the classical motion, however, the quantum $\left< v \right>$ oscillation always have smaller amplitudes since tunneling effects enhance the slipping forward of the particle.

\begin{figure}[htbp]
    \centering
    \includegraphics[width = 0.96 \textwidth]{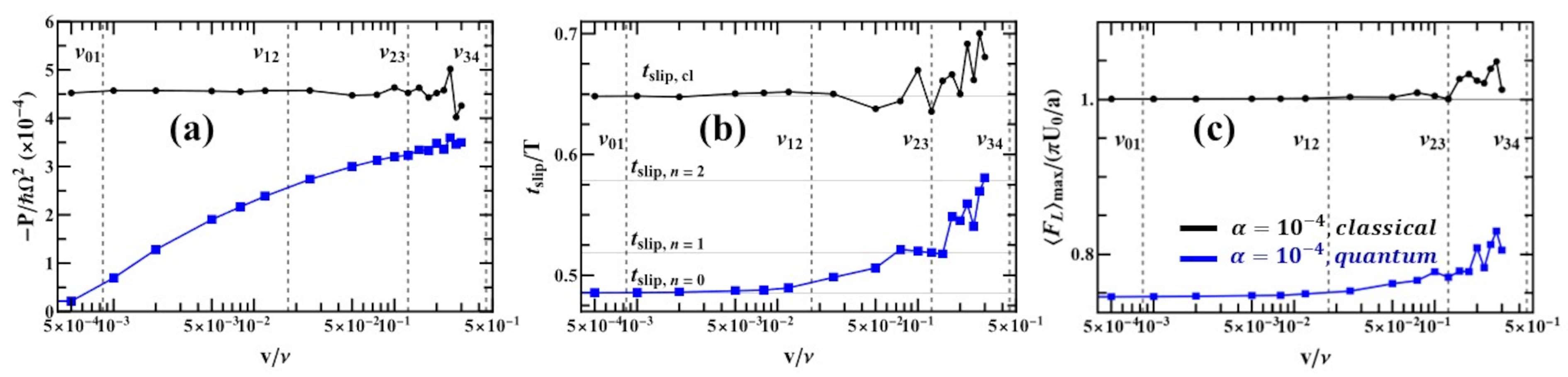}
    \caption{\label{fig:Figure4} Dynamic properties as a function of $v/\nu$ calculated in the first period of the motion. (a) Released to the external bath power $-P$ normalized by $\hbar\Omega^2$; (b) The slipping time $t_{slip}/T$ for a maximal lateral force; (c) The maximal lateral force $\left< F_L \right>_{max}$. Classical (in black) and quantum mechanical (in blue) results are given. The diabatic tunneling velocities $v_{ij}$ between the different instantaneous eigenstates are also shown.}
\end{figure}
 
Further insight into the 
$\left<x\right>$ and $\left<v\right>$ evolution can be gained by examining the time-dependence of their uncertainty principle. We find that $\sigma_x \sigma_p=\hbar/2$ during the particle sticking period $t/T < 0.5$ since the particle is confined in the ground state in the first well of the potential surface. For all later times $\sigma_x \sigma_v>\hbar/2$ and it also shows oscillatory-like behavior. For the closed classical system $\sigma_x \sigma_v=0$ as expected, while for the open classical system $\sigma_x \sigma_v > 0$ at around $t/T \approx \frac{a^2}{(2\pi)^3 \hbar \alpha}$.

We are now in a position to examine the evolution of the lateral force of the nanoparticle, {typically measured via atomic force or scanning tunneling microscopy (AFM or STM) \cite{ Socoliuc2004, Bartels1997, Liu2015, Almeida2016, Yang2018}} or optical means \cite{Erik2007,Gangloff2015,Counts2017}. The computed lateral force $\left< F_L \right>$ is defined as the spring force that holds the nanoparticle in the harmonic potential corrected by the motion of the optical trap, see Tab. \ref{tab:1}. The lateral force is a measure of the frictional force as the optical trap drags the particle above the atomic chain. The results shown in Fig.\ref{fig:Figure3}(m,n,o,p) help us delineate quantum mechanical vs classical effects for different driving velocities. As the particle starts slipping at {$t/T\sim 0.5$} it catches up with the velocity of the driving trap and  $\left< F_L \right>$ reaches its maximum. For small velocity, as mentioned above, the classical slipping occurs at the moment when the minimum of the first well of the potential surface diminishes, which happens at $t_{slip}/T \approx \frac{1}{4} + \frac{\eta}{2\pi} \approx 0.65$.  The maximum classical lateral force is $\left< F_L \right>_{max} = \dfrac{\pi U_0}{a}$. The quantum motion, on the other hand, pushes the slipping forward by $25\%$ at $t_{slip}/T \approx 0.49$ and the maximum quantum lateral force is smaller than classical one by $25\%$ i.e. $\left< F_L \right>_{max} \approx 0.75 \left( \dfrac{\pi U_0}{a} \right)$. For larger driving velocities, the quantum slipping happens at a later time, thus the maximum lateral force is larger than the one for smaller velocities.

In addition to the lateral force, the energy transfer between the moving particle and the external bath is also related to the frictional process. The transferred energy is expressed in terms of released to the environment heat $(-Q)$ defined as the difference between the energy of the particle and the work done by the moving trap (see Table \ref{tab:1}). Fig. \ref{fig:Figure3}(q-t) shows that during the stick-phase the nanoparticle does not exchange heat with the environment meaning that this regime is essentially friction-free. As the particle continues into its slip phase, higher energy levels are occupied. This excess above the ground state energy is being released to the environment marking the onset of dissipation. Comparing to the classical slipping, the quantum slipping generally occurs earlier, and as a result there is less dissipation since the quantum released heat is smaller than the classical one.

Let us now examine the velocity functional dependence of the dissipative motion of the nanoparticle. For this purpose, we track the power $P$ associated with the released to the environment heat at the end of the first period and the maximum lateral force 
 $\left< F_L \right>_{max}$ within that time frame. The results shown in Fig. \ref{fig:Figure4}(a,c) are presented together with the time $t_{slip}$ in Fig. \ref{fig:Figure4}b characterizing the moment at which the lateral force becomes maximum in the first period of the motion for each velocity. Since the maximum lateral force signals the stick-slip transition, the times $t_{slip,n}$ at which the different quantum {eigenstates} participate in this process are also shown in Fig. \ref{fig:Figure4}b. While the quantum particle dynamics is determined by a linear combination of the different eigenstates admixtures due to occurring Landau-Zener tunnelings, the classical particle with relatively slow velocity always stays at the minimum of the first potential \cite{Socoliuc2004, Gnecco2012} well until $t_{slip, cl}/T \approx \frac{1}{4} + \frac{\eta}{2\pi} \approx 0.647$. For higher velocities, higher states of the Hamiltonian become important and in both, quantum and classical limits, the maximal lateral force, released heat, and slipping time  can experience oscillatory-like features, as shown in Fig. \ref{fig:Figure4}.
 
 The heat power in Fig. \ref{fig:Figure4}a shows a notable difference between the classical and quantum motion. Since the classical particle is stuck in the first minimum of the potential, the classical $P$ is practically a constant over a large velocity range. For the quantum mechanical case, increasing the velocity causes an admixture of higher level eigenstates which results in an increasing released heat rate. Nevertheless, the quantum thermal power reaches about  $75\%$ of the classical value at $v \geq v_{23}$. A similar difference in magnitude is also found when comparing the classical and quantum maximum lateral force, as shown in Fig. \ref{fig:Figure4}c. We find that the classical $\left< F_L \right>_{max} \approx \frac{\pi U_0}{a}$ and it remains a constant over a large velocity range. The quantum $\left< F_L \right>_{max}$ is also a constant ($75\%$ of the classical one) although in a somewhat smaller velocity range.

\section{Conclusions}

Friction at the nanoscale is a complex process, and here we offer a comprehensive study delineating quantum mechanical vs classical effects in the Prandtl-Tomlinson model, one of the most popular models in tribology. {Our study examines in depth the frictional motion in several kinematic and dynamical properties. The detailed analysis of the power and frictional force, the most experimentally relevant quantity, in terms of various regimes and characteristic parameters goes beyond the stick-slip dissipation rate and its lubricity discussed in Ref. \cite{Zanca2018}. In particular, the connection  with characteristic times brings deeper understanding of the force and power dependence upon velocity beyond the adiabatic regime of motion. }
{For the stick-slip regime,  controlled by the corrugation parameter, the Landau-Zener tunneling at multiple eigenenergy level anti-crossings is responsible for allowing the particle to permeate through the potential barriers, which are practically impenetrable for the classical particle. As a result, the stick-slip quantum mechanical motion experiences reduced friction as compared to the classical case. The newly found force-velocity and power-velocity correlations are experimentally relevant relations, in which quantum vs classical regimes are clearly seen. }The quantum mechanical effects in the nanoscale friction are expected to play a reduced role at higher temperatures. As $T$ is increased, higher states will become important leading to more pronounced classical features in the motion of the particle.

The theoretical understanding is made possible by presenting the full description of the quantum mechanical and classical {methodologies} of Prandtl-Tomlinson model by distinguishing quantum vs classical vs dissipative aspects of the {nanoscale} friction. A variety of properties are defined and examined in order to track the time dynamics and velocity functionality of the frictional process. Dissipative heat, dissipative thermal power, and lateral force can serve as a theoretical base for possible experimental probing of quantum effects in stick-slip motion. Moving cold atoms and ions in optical lattices and time-resolved {Landau-Zener} tunneling in periodic potentials, already achieved in the laboratory \cite{Gangloff2015,Counts2017,Bonvin2024,Zenesini2009}, may be a useful platform to demonstrate the reduction of friction in the quantum mechanical regime.

\section{Acknowledgments}
L.M.W. acknowledges financial support from the US National Science Foundation under grant No. 2306203. P. R.-L. acknowledges support from Ministerio de Ciencia e Innovaci\'on (Spain), Agencia Estatal de Investigaci\'on, under project NAUTILUS (PID2022-139524NB-I00).

\end{document}